\documentclass[12pt,a4paper]{article}     

 \newcommand{\be}{\begin{equation}}
  \newcommand{\bea}{\begin{eqnarray}}
\newcommand{\eea}{\end{eqnarray}}
  \newcommand{\ee}{\end{equation}}
 \newcommand{\vex}{\mbox{\boldmath${\rm x}$}}
 \newcommand{\veO}{\mbox{\boldmath${\rm 0}$}}

\def\fun#1#2{\lower3.6pt\vbox{\baselineskip0pt\lineskip.9pt
\ialign{$\mathsurround=0pt#1\hfil##\hfil$\crcr#2\crcr\sim\crcr}}}

\title{Phase structure of the QCD vacuum in a magnetic
field at low temperature \\ } \author{N.O.Agasian\\ Institute of Theoretical
and Experimental Physics, \\Moscow 117218, Russia\\
e-mail: agasyan@heron.itep.ru}

\date{}

\usepackage{epsf}

\begin{document}
\maketitle
\begin{abstract}

We study the QCD phase structure in magnetic field
 $H$  at low temperature $T$. The hadronic phase free energy in a
 constant homogeneous  magnetic field is calculated in
 one-loop approximation of the chiral perturbation theory. The
 dependence of the quark and gluon condensates upon the temperature
 and field strength is found. It is shown that the chiral phase
 transition order parameter $\langle \bar q  q\rangle$  remains
 constant provided field strength and temperature are related via
 $H=const \cdot T^2$.
 \end{abstract}
 \vspace{2cm}


\newpage

 1. The investigation of the vacuum state behavior under the
 influence of the various external factors is known to be one of the
 central problems in quantum field theory. In the realm of strong
 interactions (QCD) the main factors are the temperature and
 the baryon density. At
temperatures below the chiral phase transition, $T<T_c$, the dynamics
of the system is characterized by confinement and spontaneous
breaking of chiral symmetry (SBCS). At low temperatures, $T<T_c$,
the partition function of the system is dominated by the contribution
 of the lightest  particles in the physical spectrum. In QCD this
role is played by the $\pi $ -- meson which is the Goldstone
excitation mode in chiral condensate. Therefore the low
temperature physics (the hadron phase) enables an adequate
description in terms of the effective chiral theory \cite{1,2,3}.
A very important problem is the behavior of the order parameter
(the quark condensate $\langle \bar q q\rangle)$ with the increase
of the temperature.  In the ideal gas approximation the
contribution of the thermal pions into the quark condensate is
proportional to $ T^2$
 \cite{4,5}. In chiral perturbation theory (ChPT) the two -- and
three--loops contributions ($\sim T^4$  and  $\sim T^6$
 correspondingly) into $\langle \bar q q \rangle$ have been found in
\cite{5}  and \cite{6}.

The situation  with the gluon condensate
 $\langle G^2\rangle\equiv \langle
(G^a_{\mu\nu})^2\rangle$is very different. The gluon condensate
is not an order parameter in phase transition and it does not lead
to any spontaneous symmetry breaking (SSB). At the quantum level
the trace anomaly leads to the breaking of the scale invariance
and  this in turn results in nonzero value of
 $\langle G^2 \rangle$.  However, this is not a SSB phenomenon and
hence does not lead to the appearance of the Goldstone particle. The
mass of the lowest excitation (dilaton)  is directly connected to the
gluon condensate, $ m_D\propto
 (\langle G^2\rangle)^{1/4}$.
Thus the thermal excitations of glueballs are exponentially
suppressed by the Boltzmann factor $\sim \exp
\{-m_{gl}/T\}$ and their contribution to the shift of the gluon
condensate is small ($\Delta\langle G^2\rangle/\langle G^2\rangle\sim
$0.1 \% at $T=$100 MeV) \cite{7.a}.  Next we note that in the
one-loop approximation ChPT pions are described as a gas of
massless noninteracting particles. Such a system is obviously
scale-invariant and therefore does not contribute into the trace of
the  energy-momentum tensor and correspondingly into $\langle
G^2\rangle$.  As it has been demonstrated in \cite{7} the gluon
condensate temperature dependence arises only at the ChPT three--loop
level.

Another interesting problem is the study of the vacuum phase
structure under the influence of the external magnetic field $H$.
Quarks play an active role in shaping the QCD vacuum structure.
Being dual carriers of both 'color' and 'electric' charges they
also respond to externally applied electromagnetic fields. The
vacuum of strong interactions influences some QED  processes has
been discussed in  Ref. \cite{Raf}. The behavior of the quark
condensate in the presence of a magnetic field was studied in
Nambu-Iona-Lasinio model earlier \cite{0.8}. For QCD, the
analogous investigation in the one-loop approximations was done in
\cite{8}. It was found that the quark condensate grows with the
increase of the magnetic field $H$ in both cases. It implies that
a naive analogy with superconductivity, where the order parameter
vanishes at same critical field, is not valid here. The behavior
of the gluon condensate $\langle G^2\rangle$ in the Abelian
magnetic field is also a nontrivial effect. Gluons do not carry
electric charge; nevertheless, virtual quarks produced by them
interact with electromagnetic field and lead to  the changes in
the gluon condensate.  This phenomenon was studied in \cite{10},
\cite{11} based on the low-energy theorems in QCD. The vacuum
energy density,
 the values of  $\langle G^2\rangle$  and $\langle \bar q q\rangle$
as functions of $H$ have been found in the two--loop approximation
 ChPT in \cite{11}.

The low-energy theorems, playing an important role in the
understanding of the vacuum state properties in quantum field
theory, were discovered almost at the same time as quantum field
methods appeared in particle physics (see, for example Low
theorems \cite{12}).In QCD, these theorems were obtained in the
beginning of eighties \cite{13}.  These theorems, being derived
from the very general symmetrical considerations and not depending
on the details
 of confinement mechanism, sometimes give information which is not
easy to obtain in another way. Also, they can be used as
"physically sensible" restrictions in the constructing of
effective theories. An important step  was made in \cite{14},
where low-energy theorems for
 gluodynamics were generalized to finite temperature case.

In the present paper the vacuum free energy in magnetic field at
finite temperature is calculated in the framework of ChPT. The
general relations are established which allow to obtain the
dependence of the quark and gluon condensates on $T$ and $H$. A new
phenomenon is displayed, namely the "freezing" of the chiral phase
transition order parameter by the  magnetic field when the
temperature increases. The physical meaning of this fact  is
discussed.

 2. The QCD Euclidean partition function in  external Abelian field $A_\mu$
 has the following form ($T=1/\beta$)
 \be Z=exp \left \{ -\frac{1}{4e^2}
\int^\beta_0 dx_4\int_V d^3x F^2_{\mu\nu} \right \}
 \int[DB][D\bar q][Dq]
\exp \left \{ -\int^\beta_0 dx_4\int_V d^3x {\cal L} \right \}.
\label{1}
 \ee
  Here the QCD Lagrangian in the background field is \be
 {\cal L}=\frac{1}{4g^2_0}
 (G^a_{\mu\nu})^2+ \sum_{q=u,d} \bar q[\gamma_\mu
 (\partial_\mu-iQ_q A_\mu-i\frac{\lambda^a}{2} B^a_\mu)+m_q]q,
 \label{2}
  \ee
where
    $Q_q$ -- is the matrix of the quark charges for the quarks
     $q=(u,d)$, and for the simplicity the ghost terms have been
    omitted. The free energy density is given by the relation $ \beta VF$
    $(T,H, m_q)=-\ln Z$. In the chiral limit $m_q\to 0$ Eq. (\ref{1}) yields
     the following expressions for the quark and gluon condensates \be
    \langle \bar q q\rangle (T, H)=\frac{\partial F(H,T,
    m_q)}{\partial m_q}\left \vert_{m_q=0} \right. ,
     \label{3}
      \ee
    \be
    \langle G^2\rangle (T, H)=4\frac{\partial F(H,T,
 m_q)}{\partial(1/g^2_0)}\left \vert_{m_q=0} \right.~.
 \label{4}
 \ee
The phenomenon of dimensional transmutation results in the
appearance of a nonperturbative dimensional parameter
 \be
  \Lambda= M \exp
 \left \{ \int^\infty_{\alpha_s(M)}\frac{d\alpha_s}{\beta(\alpha_s)}
 \right \}~,
  \label{5}
  \ee
  where $M$  is the ultraviolet cutoff,
   $\alpha_s=g^2_0/4\pi$, and $\beta(\alpha_s)=d\alpha_s(M)/d
  ~ln M$  is the Gell-Mann-Low  function. In chiral limit
   $(m_q=0)$ the system described by the partition function
   (\ref{1}) is characterized by three dimensionful parameters
    $M, T$ and $H$.  The free energy density is renorm-invariant
  quantity and hence its anomalous dimension is zero. Thus
   $F$ has only a normal (canonical) dimension equal to 4. Making use
  of the renorm-invariance of
   $\Lambda$,
  one can write in the most general form
  \be F=\Lambda^4 f \left (\frac{T}{\Lambda}, \frac{H}{\Lambda^2}
  \right ),
  \label{6}
   \ee
   where the function  $f$  is still unknown. From
   (\ref{5})  and  (\ref{6}) one gets
   \be
    \frac{\partial F}{\partial (1/g^2_0)}=
   \frac{\partial F}{\partial \Lambda^2}
\frac{\partial \Lambda^2}{\partial(1/g^2_0)}=
  \frac{4\pi\alpha^2_s}{\beta(\alpha_s)} (4-T\frac{\partial}{\partial
  T} -2 H\frac{\partial}{\partial H}) F.
   \label{7}
    \ee
    With the account of
  (\ref{4}) the gluon condensate is given by
   \be
   \langle G^2\rangle (T,H)
    =\frac{16\pi\alpha^2_s}{\beta(\alpha_s)}
(4-T\frac{\partial}{\partial T}-2H\frac{\partial}{\partial H}) F(T,H)
\label{8}
 \ee
 At  $T=0$, $H=0$  we recover the well known expression for
the nonperturbative vacuum energy density in the chiral limit.
In the one-loop approximation
 $(\beta=-b_0\alpha^2_s/2\pi,~~
b_0=(11 N_c-2N_f)/3)$  it has the form
 \be
\varepsilon_v=F(0,0) = -\frac{b_0}{12 8 \pi^2} \langle G^2\rangle
\label{9} \ee

 Let us now derive the low-energy theorems at finite temperature in
the presence of magnetic field. Strictly speaking,
 $\beta$-function depends on $H$ and the low-energy
  theorems acquire electromagnetic corrections
   $\propto e^4$. However, since the free energy is independent of
  the scale $M$ at which the ultraviolet divergences are regulated,
  we can choose
 $M^2\gg H, T^2, \Lambda^2$.  Hence, we can take the lowest order
  expression for
  $\beta$-function  ($\beta(\alpha_s)=-b_0\alpha_s^2/2\pi)$  and the
  electromagnetic corrections vanish. Let us introduce the field
  $\sigma (\tau=x_4, \vex)$  and operator  $\hat D$,
   \be
  \sigma (\tau,\vex) = -\frac{b_0}{32
  \pi^2} (G^a_{\mu\nu} (\tau,\vex))^2, \label{10.a} \ee
    \be \hat
 D=4-T\frac{\partial}{\partial T} -2H\frac{\partial}{\partial H}.
 \label{11.a} \ee
  Differentiating  (\ref{4}) $n$ times with respect to
 $(1/g^2_0)$ and taking into account  (\ref{7}),
 (\ref{10.a}) and (\ref{11.a}) one obtains
  $$ \hat D^{n+1}
 F=\hat D^n\langle \sigma(0,\veO)\rangle $$ \be =\int
 d\tau_nd^3 x_n...  \int d\tau_1d^3 x_1 \langle \sigma(\tau_n,
\vex_n)...  \sigma(\tau_1, \vex_1) \sigma(0, \veO)\rangle_c.
\label{12.a} \ee The subscript  $c$ means that only connected
diagrams are to be included. Proceeding in the same way, it is
possible to obtain the similar theorems for renorm-invariant
operator
 $O(x)$ which is built from quark and/or gluon fields
 $$ \left( T \frac{\partial}{\partial
T} + 2 H \frac{\partial}{\partial H} -d\right )^n\langle O\rangle
$$ \be =\int d\tau_n d^3 x_n...  \int d\tau_1 d^3 x_1 \langle
\sigma(\tau_n, \vex_n)...  \sigma(\tau_1, \vex_1) O(0,
\veO)\rangle_c.  \label{13.a} \ee where  $d$ is the canonical
dimension of operator
 $O$.  If operator  $O$  has also anomalous dimension, the
appropriate
 $\gamma$-function should be included.

3.The above equations enable to obtain the values of the
condensates as functions of
 $T$  and  $H$ provided the free energy density is known. To get the
latter the ChPT will be used. At low temperatures
 $T<T_c (T_c$ is the chiral phase transition temperature) and for weak
fields
 $H< \mu^2_{hadr} \sim (4\pi F_\pi)^2$ the characteristic momenta
in the vacuum loops are small and theory is adequately described
by the
        low-energy effective
 chiral Lagrangian
 $L_{eff}$ \cite{2,3}.
This Lagrangian can be represented as a series expansion  over
the momenta (derivatives) and quark masses
 \be
L_{eff}=L^{(2)}+L^{(4)}+L^{(6)}+...
\label{10}
 \ee
 The leading term in
(\ref{10}) is similar to the Lagrangian of the non-linear sigma model
in the external field
 $$ L^{(2)}=\frac{F^2_\pi}{4}Tr(\nabla_\mu
U^+\nabla_\mu U)+\Sigma Re~Tr(\hat M U^+), $$ \be \nabla_\mu
U=\partial_\mu U-i[U,V_\mu].
 \label{11}
  \ee
  Here $U$ is a unitary
$SU(2)$  matrix, $F_\pi=93 $MeV  is the pion decay constant, and
 $\Sigma$  has the meaning of the quark condensate
 $\Sigma =|\langle \bar u u\rangle | = |\langle
\bar d d\rangle |$.  The external Abelian magnetic field $H$ is aligned
 along the  $z$ -axis and corresponds to $V_\mu(x)=(\tau^3/2) A_\mu(x)$
 with the vector-potential $A_\mu$  chosen as $A_\mu(x)=\delta_{\mu 2}Hx_1$.
The mass difference between the
 $u$ and  $d$ quarks appears in the effective chiral Lagrangian
 only quadratically. Further, to obtain an expression for the
quark condensate in the chiral limit we use only the first
 derivative with respect to the mass of one of the quarks.
Therefore, we can neglect the mass difference between the $u$ and $d$
quarks and assume the mass matrix to be diagonal $\hat M=m\hat I$.

At  $T<T_c, H< \mu^2_{hadr}$ the QCD partition function coincides
with the partition function of the effective chiral theory
 \be
  Z_{eff}[T,H]=e^{-\beta
VF_{eff}[T,H]}=Z_0[H]\int [DU] \exp \{ -\int^\beta_0 dx_4\int_Vd^3x
L_{eff} [U,A]\}
 \label{12}
  \ee
  At the one-loop level it is sufficient to restrict the expansion
of $L_{eff}$  by the quadratic terms with respect to the pion
field. Using the exponential parameterization of the matrix $U(x)=
\exp \{
 i\tau^a\pi^a(x) /F_\pi\} $ one finds \be L^{(2)}=\frac12 (\partial_\mu
\pi^0)+\frac12 M^2_\pi(\pi^0)^2+ (\partial_\mu\pi^++iA_\mu\pi^+)
(\partial_\mu\pi^--iA_\mu\pi^-)   + M^2_\pi \pi^+\pi^-, \label{13} \ee where
 the charged $\pi^\pm$  and neutral  $\pi^0$  meson fields are introduced
 \be
\pi^\pm =(\pi^1\pm i\pi^2)/\sqrt{2},~~ \pi^0=\pi^3
 \label{14}
  \ee
Thus  (\ref{12})  can be recasted into the form\footnote{ The
partition function $Z_{eff}^R$ describes charged $\pi^{\pm}$ and
neutral $\pi^0$ ideal Bose gas in magnetic field. Relativistic
charged Bose gas in magnetic field at finite temperature and
density with application to Bose-Einstein condensation and Meissner
effect was studied in Refs. \cite{Dai}, \cite{Tom}, \cite{Elm},
\cite{Roj}.} \be Z_{eff}^R[T,H]=Z^{-1}_{p.t.} Z_0[H]\int [D\pi^0]
[D\pi^+][D\pi^-] \exp \{-\int^\beta_0 dx_4\int_Vd^3x L^{(2)}[\pi,
A]\} \label{15} \ee
    where partition  function is normalized for the case of
    perturbation theory  at
     $T=0, H=0$ \be Z_{p.t.} = [\det
    (-\partial^2_\mu+M^2_\pi)]^{-3/2}.  \label{16.a}
     \ee
     Integration of
    (\ref{15}) over $\pi$-fields leads to
     \be
     Z_{eff}^R=Z^{-1}_{p.t.}Z_0[H] [{{\rm
 det}_T(-\partial^2_\mu+M^2_\pi)}]^{-1/2} [{{\rm
 det}_T(-|D_\mu|^2+M^2_\pi)}]^{-1}, \label{17.a} \ee
 where
 $D_\mu=\partial_\mu-iA_\mu$ is a covariant derivative
 and a symbol  $T$  means that the determinant is calculated
 at finite temperature
  $T$ according to standard Matsubara rules. Taking
  (\ref{16.a})  into account and regrouping multipliers in
  (\ref{17.a}) one gets the following expression for $Z_{eff}^R$
  $$ Z_{eff}^R [T,H]=Z_0[H] \left[
 \frac{\det_T(-\partial^2_\mu+M^2_\pi)}
   {\det(-\partial^2_\mu+M^2_\pi)}\right ] ^{-1/2}
  \left[
   \frac{\det
 (-|D_\mu|^2+M^2_\pi)}
   {\det(-\partial^2_\mu+M^2_\pi)}\right ] ^{-1}
  $$
 \be
 \times \left[
   \frac{\det_T
 (-|D_\mu|^2+M^2_\pi)}
   {\det(-|D_\mu|^2+M^2_\pi}\right ] ^{-1}
 \label{18.a}
 \ee
 Then the effective free energy can be written in the form
  \be F_{eff}^R(T,
 H)=-\frac{1}{\beta V}\ln Z_{eff}^R
 =\frac{H^2}{2e^2}+F_{\pi^0}(T)+F_{\pi^{\pm}}(H) +F_s(T,H).
 \label{16} \ee
 Here  $F_{\pi^0}$  is the free energy of massive scalar boson
 \be
F_{\pi^0}(T)=T\int\frac{d^3p}{(2\pi)^3}\ln (1-\exp (-\sqrt{{\bf
p}^{2}+M^2_\pi}/T)),
 \label{17}
 \ee
 $F_{\pi^\pm}$ is a Schwinger result for the vacuum energy
density of charged scalar particles in the magnetic field.
 \be
F_{\pi^\pm}(H)=-\frac{1}{16\pi^2}\int^\infty_0\frac{ds}{s^3}
e^{-M^2_\pi s}  [\frac{Hs}{\sinh (Hs)}-1],
\label{18}
\ee
and
$$
F_s(T,H)=\frac{HT}{\pi^2}\sum^\infty_{n=0}\int^\infty_0 dk\ln
(1-\exp(-\omega_n/T)),
$$
\be \omega_n=\sqrt{k^2+M^2_\pi+H(2n+1)}, \label{19} \ee where
$\omega_n$ are Landau levels of the  $\pi^\pm$ mesons in constant
field $H$. \footnote{Technically, a transition
 for the free energy
  $F=\frac12 Tr \ln
 (p^2_4+\omega^2_0(\bf p))$
 from the vacuum case ($H=0, T=0)$ to the case of  $ H\neq 0, T\neq
 0$
 is straightforward.
  Omitting the details of
  the calculations, we note that, eventually, this
transition reduces to the substitutions
  $p_4\to \omega_k=2\pi kT$
    ($k=0,\pm
 1,...$), $\omega_0=\sqrt{{\bf p}^2+M^2_\pi}
 \to \omega_n=\sqrt{p^2_z+M^2_\pi+H(2n+1)}
 $
  and
 $Tr\to \frac{HT}{2\pi} \sum^\infty_{n=0}
\sum^{+\infty}_{k=-\infty}\int^{+\infty}_{-\infty}
\frac{dp_z}{2\pi},$
 where the degeneracy multiplicity of  $H/2\pi$ has been
 taken into
 account for the Landau levels. Performing summation over Matsubara
 frequencies, we obtain (26).}

4.The free energy  $F_{eff}^R$  determines the thermodynamical
properties and the phase structure of the QCD vacuum state below the
temperature of the chiral phase transition, i.e. in the phase  of
confinement.

 Equations
 (\ref{8}) and  (\ref{16})  describe the dependence of
  $\langle G^2\rangle$  on  $T$  and  $H$.  The action of the
operator $\hat D$  on  $F_{eff}^R$  leads to  $\hat
D F_{\pi^0}(T)=0$ since  $M^2_\pi\to 0$ and $
F_{\pi^0}(T)\sim T^4$   in chiral limit and thus
(4--$T\partial/\partial T)F_{\pi^0}(T)=0$. It can be easily shown by
direct calculation that
 $\hat D F_s(T, H)=0$. The nontrivial dependence of
 $\langle G^2\rangle$  on $H$  arises only due to Schwinger
term  $F_{\pi^\pm}(H)$
\be \langle G^2\rangle (T, H)= \langle G^2\rangle
+\frac{\alpha^2_s}{3\pi\beta(\alpha_s)} H^2
\label{20}
\ee
Next we note that because of the asymptotic freedom the QCD
$\beta$-function, $\beta(\alpha_s)=-b_0\alpha_s^2/2\pi+...$ is negative and
 hence the gluon condensate diminishes with the $H$ increasing
$$ \langle G^2\rangle (T,H)=\langle G^2 \rangle -\frac{2}{3b_0}
H^2.$$
Thus, the temperature corrections to the gluon condensate
in magnetic field vanish at the ChPT one-loop level.

 In order to get the dependence of the quark condensate upon
  $T$ and  $H$
 use is made of the Gell-Mann-Oakes--Renner relation (GMOR)
  \be F^2_\pi
 M^2_\pi=-\frac12(m_u+m_d)\langle\bar u u +\bar d d\rangle =2m \Sigma
 \label{21} \ee
 Substituting (\ref{16}) into (\ref{3}), calculating the derivative
 over $M^2_\pi$ and then taking the limit $M^2_\pi\to 0$ one gets
 $$ \langle \bar q q\rangle (T,H)= \langle \bar q
 q\rangle (1-\frac13\cdot \frac{T^2}{8F^2_\pi}+\frac{H}{(4\pi
 F_\pi)^2} \ln 2-\frac{H}{2\pi^2F^2_\pi} \varphi(\frac{\sqrt{H}}{T}))
 $$ \be
\varphi(\lambda)=\sum^\infty_{n=0}\int^\infty_0\frac{dx}{\omega_n(x)
(\exp(\lambda\omega_n(x))-1)}, ~~ \omega_n(x)=\sqrt{x^2+2n+1}
\label{22} \ee Now we consider various limiting cases. In the
strong  field, $\sqrt{H} \gg T$ $(\lambda \gg 1)$, the lowest
Landau level ($n=0$) gives the main contribution to the sum
(\ref{22})
\be
\varphi(\lambda\gg 1) = \sqrt{\frac{\pi}{2\lambda}}e^{-\lambda}
  +O(e^{-\sqrt{3}\lambda}).
  \label{23}
  \ee
   In the opposite limit of weak field, $\sqrt{H}\ll T(\lambda\ll 1)$, the
  sum in (\ref{22}) is calculated with required accuracy using
  the Euler-MacLaren formula. Furthermore, one gets the following result with the use of
  the  asymptotic expansion of integral (\ref{22})\cite{Kap} at $\lambda\ll 1$
  \be
  \varphi (\lambda\ll 1) =\frac{\pi^2}{6} \frac{1}{\lambda^2}
  +\frac{7\pi}{24} \frac{1}{\lambda} +\frac{1}{4} \ln \lambda
  +C+\frac{ \zeta(3)}{48\pi^2} \lambda^2 +O(\lambda^4),
  \label{24}
  \ee
   here $C=\frac14(\gamma-\ln 4\pi-\frac16),~~ \gamma =0.577...$
   is Euler's constant and $\zeta(3)=1.202$ is Riemann zeta function.
  Thus, one obtains the following limiting expressions for the quark condensate in the chiral limit in a magnetic
  field at $T\neq 0$
  \be
  \frac{\langle \bar qq\rangle (T,H)}{\langle \bar q
  q\rangle}=1-\frac13\cdot \frac{T^2}{8F^2_\pi}
  +\frac{H}{(4\pi F_\pi)^2} \ln2-\frac{H^{3/4}T^{1/2}}{(2\pi)^{3/2}
  F^2_\pi} e^{-\sqrt{H}/T},~~\sqrt{H}\gg T
  \label{25}
  \ee
  and
  \be
  \frac{\langle \bar qq\rangle (T,H)}{\langle \bar q
  q\rangle}=1-\frac{T^2}{8F^2_\pi}
  +\frac{H}{(4\pi F_\pi)^2}
  A-\frac{7 \sqrt{H}T}{48 \pi F_\pi^2}
  - \frac{H}{(4\pi F_\pi)^2}
  \ln\frac {H}{T^2},
  ~~\sqrt{H}\ll T \label{26} \ee
  where
  $A=\ln 2+8C\simeq 4.93$.

   In the framework of ChPT the quark condensate (\ref{22}) at
   $H\neq 0$, $T\neq 0$ is determined by three dimensionless parameters
   $H/(4\pi F_{\pi})^2$, $T^2/F_{\pi}^2$ and $\lambda=\sqrt{H}/T$.
   The quantity $\lambda$ is a natural dimensionless parameter in
   this approach. The motion of a particle (massless pion) in the
   field is characterized by the curvature radius of it's
   trajectory, and in the magnetic field this is the Larmor radius
   $R_L=1/\sqrt{H}$. On the other hand, there is another length $l_T=1/T$ -
   "temperature length" at $T\neq 0$.
   Therefore, charged $\pi^{\pm}$-mesons in
   magnetic field effectively acquire "mass", $m_{eff} = \sqrt{H}$,
   determined by the lowest Landau level, when Larmor radius of a
   particle in the field is much less than $l_T (\lambda \gg 1)$.
   Correspondingly, their
   contribution to the shift of the chiral condensate is
   suppressed by the Boltzman factor $\propto\exp\{-m_{eff} / T\}$.
   In the weak field limit $\pi^{\pm}$-mesons give
   standard temperature one-loop approximation ChPT contribution
   to $\langle \bar q q\rangle$. Besides, additional temperature
   and magnetic corrections appear.
   Neutral $\pi^0$-meson contributes to
   $\langle \bar q q\rangle(T,H)$ as usual massless scalar
   particle.

   An interesting
  phenomenon reveals itself in the vacuum QCD phase structure  under
  consideration. One can find from (\ref{22}) such a function $H(T)$
  that the chiral condensate $\langle \bar q q\rangle ( T,H)$ remains
  unchanged when the temperature and magnetic field change in
  accordance with $H_*=H(T)$. Then $H_*$ is found by solving the
  following equation (see (\ref{22}) $\langle \bar q q\rangle
  (T,H_*)- \langle \bar q q\rangle=0)$
   \be 1-\frac{3}{2\pi^2} \lambda^2\ln 2
  +\frac{ 12}{\pi^2}\lambda^2\varphi(\lambda)=0 \label{27} \ee
  The numerical
  solution of  (\ref{27})  yields $\lambda_*=0.111...$ Thus, quark
  condensate stays unchanged when $T$ and $H$ are increased according
  to $H=0.013\cdot T^2$.  Hence it is possible to say that the order
  parameter $\langle \bar q q\rangle$  of the chiral phase transition
  is "frozen" by the magnetic field.

\begin{figure}[h]
\epsfysize=5cm
\epsfbox{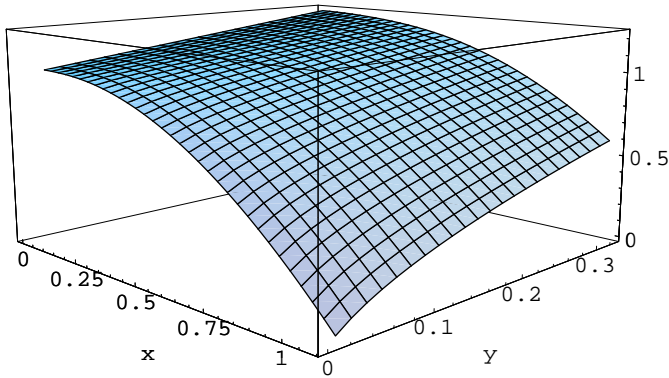}
\caption{Quark condensate  $\Sigma (T,H)/ \Sigma$ as function of
 $x=T/\sqrt{8}F_\pi$
 and $y=H/(4\pi F_\pi)^2$}
\end{figure}

  Note that $H(T_c)/ (4\pi F_\pi)^2\simeq 2\cdot 10^{-4} \ll 1$ at $T=T_c\simeq 150 $MeV
and therefore the above relations remain valid up to the deconfined phase
transition point.  In the vicinity of $T_c$ the effective low energy chiral
Lagrangian fails to provide an adequate description of the QCD vacuum
thermodynamical properties, and strictly speaking becomes physically invalid.
  The following is worth noting. In deriving
(\ref{22}), at the first step the physical quantity
 as functions of $M_\pi$ where obtained, and
only then the chiral limit
 $M_\pi\to
0$ was taken. Acting in the inversed sequence we would have
obtained all temperature corrections to condensate identically
equal to zero. This points to the fundamental difference of the
two cases: the exactly massless particle and the particle with
infinitesimal small mass.

5. It has been shown in the present letter that the quark
condensate is "frozen" by the magnetic field when both temperature
$T$ and magnetic field $H$ are increased according to the
 $H=const \cdot T^2$ law. This points to the fact that the direct analogy
  between the quark condensate in QCD and the theory of
  superconductivity is untenable. In the BCS theory the Cooper pairs
  condensate is extinguished by the temperature and magnetic field.
  The "freezing" phenomenon can be understood in terms of the
  general Le Chatelier--Braun principle \footnote{The external action
  disturbing the system from the equilibrium state induces processes
  in this system which tend to reduce the result of this action}.The
  external field contributes into the system an additional energy
  density
  $H^2/2$. The system tends to compensate this energy change and to
  decrease the free energy by
  increasing the absolute value of the quark condensate:
  $\Delta\varepsilon_v=-m|\Sigma(H)-\Sigma(0)|<0$. On the other hand,
  if the temperature of the system is increased (by bringing some
  heat into it) the processes with heat absorbtion
  by damping the condensate are switched on. The interplay of these
  processes is at the origin of the above "freezing" of $\Sigma
  (T,H)$.  Next, since gluons do not carry electric charge, the
  magnetic field affects the gluon sector of the vacuum only
  indirectly via the quark sector and thus the  Le Chatelier--Braun
   principle is not applicable directly to the gluon condensate.  For
  the same reason gluon condensate decreases nonlinearly
  with $H$ increasing according to $\Delta \langle G^2\rangle\propto
  -H^2$, while for the quark condensate $\Delta \Sigma
  \propto H$.

The author is grateful to B.L.Ioffe, V.A.Novikov, Yu.A.Simonov,
A.V.Smilga and S.M.Fedorov for comments and discussions. The
financial support of RFFI grant 00-02-17836 is gratefully
acknowledged.


\begin{thebibliography}{99}

\bibitem{1}
S.Weinberg, Physica {\bf 96 a} (1979) 327.
\bibitem{2}
J.Gasser and H.Leutwyler, Ann. Phys. {\bf 158} (1984) 142.
\bibitem{3}
J.Gasser and H.Leutwyler, Nucl. Phys. {\bf B250} (1985) 465.
\bibitem{4}
P.Bin\'{e}trui and M.K.Gaillard, Phys. Rev. {\bf D32} (1985) 931.
\bibitem{5}
J.Gasser and H.Leutwyler,  Phys. Lett. {\bf B184} (1987) 83; {\bf
B188}  (1987) 477; H.Neuberger, Phys. Rev. Lett. {\bf 60}
(1988) 889.
\bibitem{6}
 H.Leutwyler, Nucl. Phys. {\bf B} (proc. Suppl.)  {\bf 4} (1988) 248;
 P.Gerber and H.Leutwyler, Nucl. Phys. {\bf B321 } (1989) 387.
 \bibitem{7.a}
N.O.Agasian, JETP Lett. 57 (1993) 208
  \bibitem{7}
  H.Leutwyler, Restoration of Chiral Symmetry, Lecture given  at
  Workshop on Effective Field Theories, Dobogoko, Hungary, 1991; Bern
  Univ. -BUTP-91-43.
  \bibitem{Raf} J.Rafelski, H.-Th.Elze, hep-ph/9806389;
  H.-Th.Elze, B.Muller, J.Rafelski, hep-ph/9811372.
  \bibitem{0.8} S.P.Klevansky, R.H.Lemmer, Phys. Rev. {\bf D39}
  (1989) 3478.
     \bibitem{8}
  I.A.Shushpanov and A.V.Smilga, Phys. Lett. {\bf B402} (1997) 351.
    \bibitem{10}
  N.O.Agasian, I.A.Shushpanov, JETP Lett. {\bf 70}
  (1999) 717.
    \bibitem{11}
     N.O.Agasian, I.A.Shushpanov, Phys. Lett. {\bf
  B472} (2000) 143.
 \bibitem{12}
 F.E.Low, Phys. Rev. {\bf 110} (1958) 974.
 \bibitem{13}
 V.A.Novikov, M.A.Shifman, A.I.Vainshtein, V.I.Zakharov, Nucl. Phys.
 {\bf B191} (1981) 301; Sov. J. Part. Nucl. {\bf 13} (1982) 224;
 A.A.Migdal, M.A.Shifman, Phys. Lett. {\bf B114} (1982) 445.
  \bibitem{14}
   P.J.Ellis, J.I.Kapusta, H.-B.Tang, Phys. Lett. {\bf
  B443} (1998) 63.
\bibitem{Dai}
J.Daicic, N.E.Frankel, V.Kowalenko, Phys. Rev. Lett.
{\bf 71} (1993) 1779
\bibitem{Tom}
 D.J.Toms, Phys. Lett {\bf B343} (1995) 259;
 Phys. Rev. {\bf D55} (1997) 7797;
 cond-mat/9612003.
 \bibitem{Elm}
  P.Elmfors, P.Liljenberg, D.Persson, Bo-S.Skagerstam,
      Phys, Lett. {\bf B348} (1995) 462.
  \bibitem{Roj}
 H.P.Rojas, Phys. Lett. {\bf B379} (1996) 148.
  \bibitem{Kap}
   J.I.Kapusta, Finite Temperature Field Theory (Cambridge
   University Press, Cambridge, 1989).

   \end{thebibliography}
    \end{document}